*This manuscript has been authored by UT-Battelle, LLC under Contract No. DE-AC05-00OR22725 with the U.S. Department of Energy. The United States Government retains and the publisher, by accepting the article for publication, acknowledges that the United States Government retains a non-exclusive, paid-up, irrevocable, world-wide license to publish or reproduce the published form of this manuscript, or allow others to do so, for United States Government purposes. The Department of Energy will provide public access to these results of federally sponsored research in accordance with the DOE Public Access Plan (http://energy.gov/downloads/doe-public-access-plan).*

# Simulation of lateral ion migration during electroforming process

Jingjia Meng[1], Enkui Lian[1], Jonathan D. Poplawsky[2], Marek Skowronski[1]

[1]Department of Materials Science and Engineering, Carnegie Mellon University, Pittsburgh, Pennsylvania, 15213, United States

[2] Center for Nanophase Materials Sciences, Oak Ridge National Laboratory, Oak Ridge, Tennessee 37831, United States

**ABSTRACT**

Non-volatile memory devices have received a lot of interest in both industry and academia in the last decade. Transition metal oxide-based memories offer potential applications as universal memory and artificial synapses. Here we focus on the one-time conditioning of metal / oxide / metal structures leading to the formation of a conducting filament in TiN/Ta$_x$O$_{1-x}$/TiN structures and develop a finite element model of this process. Model solved coupled equations for charge, heat, and mass transport with the last one including concentration, temperature, and stress gradients as driving forces. The results closely replicated available structural data such as diameter and composition of Ta-enriched core of the filament, O-enriched ring around it, and the dynamics of filament formation. The range of critical material parameters, namely heats of transport for Ta and O, is discussed.

# 1. INTRODUCTION

Oxide-based resistive switching devices show promise for non-volatile solid state memory applications[1] with the first commercial memory product introduced by Panasonic in 2013. They also can potentially be used in neuromorphic circuits serving as artificial synapses.[2] This potential prompted wide ranging investigations including exploration of different materials and device structures, and fundamental studies of processes involved in memory switching. While research efforts have produced a lot of specific data, we still do not have a comprehensive understanding of the switching processes in a form of quantitative physics-based device model.

Great many resistive switching device types have been fabricated, tested, characterized, and their operating mechanism investigated. Here we focus only on one class of such devices referred to as Valence Change Memories (VCM)[3] which encode information by locally changing the composition of the functional oxide by motion of ions. VCM devices switch from Low Resistance State (LRS) to High Resistance State (HRS) when the voltage applied to the terminals exceeds $V_{SET}$. Switching in opposite direction, referred to as RESET process, is induced by application of pulse of opposite polarity. The changes in conductance were interpreted as due to formation of a small diameter conducting filament within the oxide matrix.[3] Even within this class of switching devices, the proposed interpretations span a wide range of phenomena incorporated into computational models of formation and switching. They include vertical motion of ions creating a gap in the filament and corresponding HRS[4,5], appearance of a constriction in the middle of the filament producing quantum point contact effects,[6,7] lateral motion of ions,[8,9] and electric field induced exchange of oxygen ions between oxide and electrodes.[10–12] The electrical conductivity mechanisms considered included band transport, thermionic emission over Schottky barriers, trap assisted tunneling, Fowler-Nordheim tunneling, and small polaron motion. Neither one of the two lists is all inclusive as the concepts proliferate with time.

The device simulations performed to date have focused on reproducing salient features of the electrical quasi-static switching *I-V* characteristics. This was invariably successful despite different mechanisms assumed by the models used to interpret results on the same material and device structure. It is quite apparent that the quasi-static *I-V* data are not characteristic enough to allow for convergence on the unique model. The set of initial experimental data to be replicated needs to be significantly broader.

In addition to the narrow experimental base, virtually all modelling efforts to date shared several major weaknesses. The first one is an almost universal assumption of the only mobile ion specie in the functional oxide layer being the oxygen ion.[3] Scanning tunneling microscopy experiments, however, provided strong evidence that metal ions are mobile in $Ta_xO_{1-x}$, $Hf_xO_{1-x}$, and $Ti_xO_{1-x}$ and their motion contributes to resistance changes.[13] (Here and throughout this report we will use $Ta_xO_{1-x}$ notation for binary mixture with x corresponding to atomic fraction of the metal.) This result was confirmed by Electron Energy Loss Spectroscopy which imaged Ta ions moving into $HfO_2$ layer during electro-formation.[14] Similar conclusion was reached by Ma *et al*. who produced detailed elemental maps of Ta distribution using High Angle Annular Dark Field and X-ray Energy Dispersive Spectroscopy.[15–17] The maps of TiN / $Ta_xO_{1-x}$ / TiN structures revealed that Ta accumulates in the core of the filament and at the same time creates a Ta-depleted ring around it. Concomitantly, Ta ions move along the direction of the electric field creating a gap in the filament located next to the anode.[16,18] The extent of Ta ion motion appeared to be higher than that of oxygen.

Another common weakness is the assumed small range of local composition changes during electro-formation and switching. The prevailing approach was to treat the oxide as a single-phase compound with point defects (namely oxygen vacancies) in the diluted limit with composition changes below 1%.[12,19–22] This allowed for neglecting the stress effects induced by the motion of ions. We do know now that composition changes induced by electro-formation and switching in many device structures are large. Composition of the filament in $Ta_xO_{1-x}$-based devices was directly assessed by HAADF to be as metal-rich as $Ta_{0.67}O_{0.33}$ compared to the starting oxide composition of $Ta_{0.31}O_{0.69}$.[16] Similar conclusions have been reached by electrical transport measurements on formed devices and comparison of results with those obtained on as-deposited oxygen-poor films.[23,24] In order to account for large composition changes, the stress term must be included in the equation describing redistribution of ions.[25–28]

Lastly, most of the modelling efforts focused on the electric field-induced migration while neglecting the temperature gradient as a possible driving force. Mass transport induced by temperature gradient is referred to as thermodiffusion, Soret effect, or thermophoresis. It is easy to argue that thermodiffusion is playing an important role in electroformation of VCM devices and their endurance. The small diameter filament with the composition very different from that of the surrounding material exhibits large radial composition gradients. Since the filament is hot enough to allow for ion motion, this gradient would induce the Fick's diffusion flux that would homogenize the material and destroy the filament in few switching cycles. For the filament to be stable and the device to switch millions of times, there must be an ion flux counteracting the Fick's diffusion. Thermodiffusion appears to the only plausible candidate. It is also worth noting that the diameter of the filament is determined by a balance of opposite fluxes: the formation of the filament due to thermodiffusion is opposed by Fick's diffusion- and by the stress gradient-induced flux. This opens up a way to calculate the diameter of the filament.[9]

Of primary interest in this report is to demonstrate how the lateral ion fluxes lead to creation of a conducting filament in an initially uniform $Ta_{0.33}O_{0.67}$ layer sandwiched between two TiN electrodes. We have selected this particular device stack due to its simplicity and a large number of electrical, structural, and modelling data available. For example, it is now well established that the electro-formation in this structure occurs in two distinct stages as has been suggested as early as 2011.[29] The first one is the electro-thermal threshold switching event which produces a rapid increase of current and temperature in the device and a spontaneous current confinement.[30–32] The modelling of these phenomena yielded a detailed temperature distribution in the device at early stages of electro-formation. The second stage is associated with ion motion induced by the temperature and electrostatic potential gradients.[33] As is frequently the case of positive feedback runaway process, the dynamics of this stage exhibited characteristic incubation period, followed by a rapid change in ion concentration profile, and saturation at steady state.[33] In addition to electrical data, transmission electron microscopy determined the filament size, shape, and composition.[16,18] The elemental maps include all elements in the structure: Ta and O from the functional layer and Ti and N from the electrodes. Below we describe a model that can account for these observations.

2. RESULTS AND DISCUSSION

2.1 Device Structure and Model Construction.

The structure modelled in this work was a TiN/$Ta_{0.333}O_{0.667}$/TiN structure with 50 nm thick uniform functional oxide and the lateral size of 150 nm (Figure 1). The structure was integrated with an on-chip 37 kΩ load resistor in order to prevent current overshoot due to discharge of parasitic capacitances.[34] Detailed description of the device fabrication is included in Supplementary Information.

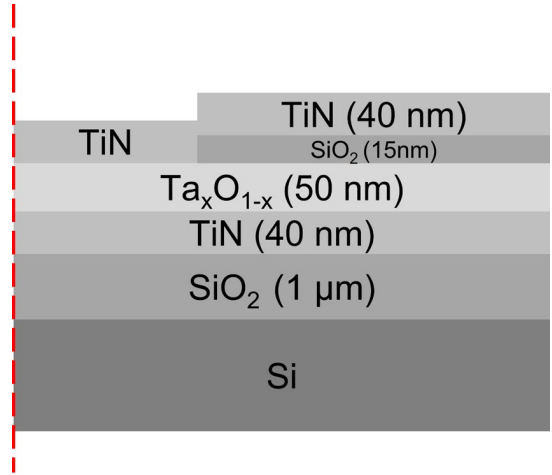

Figure 1. Geometry of inverted via memory device used in this work. Device is circular with red dashed line representing the center axis.

Table 1 in the Supplementary Information lists materials property values used in the modelling. The critical input parameter is the electrical conductivity of the functional oxide as a function of composition and temperature. We have measured the conductivity of as-deposited $Ta_xO_{1-x}$ films in the composition range of $0.299<x<0.370$ and temperatures between 300K and 600K (Figure 2(a)). The conductivity of all investigated samples increased with increasing temperature in a linear fashion when plotted as $ln\sigma$ vs. $1/T$ as is typical of semiconductors and insulators. The activation energy is decreasing with Ta content eventually reaching zero when material becomes metallic.[24]

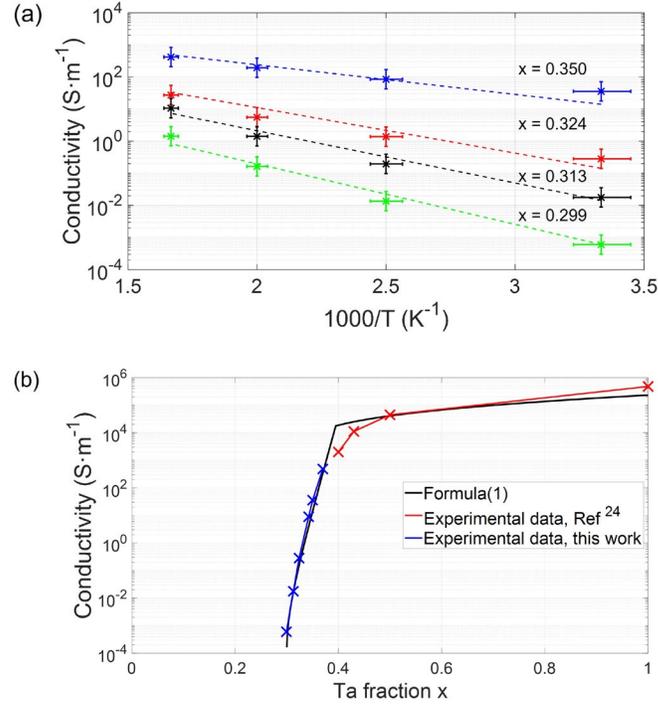

Figure 2 (a) Experimental values of electrical conductivity as a function of temperature and composition. (b) Electrical conductivity at 300K in the entire 0.286<x<1.00 composition range including data from Ref.[24].

The experimental conductivity data have been fitted with the formula:

$$\sigma(x,T) = \sigma_o(x) e^{-\frac{E_{COND}(x)}{kT}} \tag{1}$$

where $E_{COND}(x)$ is the activation energy and $\sigma_o$ is the factor depending on the density of centers responsible for generation of free carriers and their mobility. Both parameters have been extracted directly from the experiment and approximated by linear functions of composition:

$$E_{COND}(x) = \begin{cases} -3.75 \times x + 1.496 \ (eV) & 0.286 < x < 0.392 \\ -0.025 \times x + 0.025 \ (eV) & 0.392 < x < 1.00 \end{cases} \tag{2}$$

$$\sigma_o(x) = 3.30 \times 10^5 \times x - 9.97 \times 10^4 \ (eV) \qquad 0.286 < x < 1.00 \tag{3}$$

The experimental data and fitting results are shown in Figure. 2 (a). Since the filament composition range of interest is wider than our deposition system can produce, we have adopted the data by Rosario *et al.*[24] Figure 2 (b) shows the conductivity versus composition at room temperature. The blue crosses represent our experimental data, the red crosses represent the results of Rosario's experiment, and the black curve is the regression fit. It is apparent that the fitting function reproduces experimental data quite well with a slight overestimate of the conductivity at fractional concentration of Ta near 0.4.

We have modelled the charge transport in the device by solving Poisson's equation with $\sigma(x,T)$ as input by solving:

$$\nabla(\sigma\nabla\varphi) = 0 \tag{4}$$

where $\varphi$ is the electrostatic potential. In this part of the model, we have neglected the ion contribution to conductivity. Coupled with charge transport was the heat transfer equation:

$$\rho C_p \frac{\partial T}{\partial t} - \nabla \cdot (\mathrm{k_{th}}\nabla T) = \vec{J} \cdot \vec{E} \tag{5}$$

where $\rho$ is the mass density of $Ta_xO_{1-x}$, $C_p$ is the thermal capacitance of a given material, $\mathrm{k_{th}}$ is the thermal conductivity, *J* is the current density, *E* is the electric filed, *T* is the temperature, and *t* is the time. Material parameters $C_p$, $k_{th}$, and $\rho$ were assumed to be constants independent of composition and temperature of $TaO_x$. The third coupled effect is the mass transport:

$$\frac{\partial c_i}{\partial t} = \nabla \cdot \left( D_i(T) \left( \nabla c_i + \frac{S_o^i}{kT^2} c_i \nabla T - \frac{\Omega_i}{kT} c_i \nabla \sigma_{HYDRO} \right) \right) \tag{6}$$

and

$$\sigma_{HYDRO} = B \frac{c_{Ta} + c_O}{c_{Ta}^o + c_O^o} \tag{7}$$

The three terms in parenthesis on right side of eq. (6) represent fluxes induced by the concentration, temperature, and hydrostatic stress gradients. The subscript *i* corresponds to the two types of ions in the material: Ta and O. $c_i$ is the number of atoms of type *i* per unit volume, $D_i(T)$ is the diffusion coefficient, $S_o^i$ represents the heat of transport, $\Omega_i$ is the ionic volume, and *B* is the bulk modulus of the film. The parameters $D_i$, $S_o^i$, $\Omega_i$, and *B* were assumed to be independent of $Ta_xO_{1-x}$ composition and, with exception of $D_i$, of temperature. The equation does not include the term due to electromigration. Electromigration is almost entirely vertical while here we focus primarily on the mechanism of the filament formation relying on lateral motion. As the consequence, the model does not describe the switching effect which will be addressed in a separate publication. Detailed description of modelling assumptions can be found in Supplementary Information.

The data on diffusivities of Ta and O ions in amorphous $Ta_xO_{1-x}$ to be used in eq. (6) are scarce. It is clear that both Ta and O ions are mobile and according to molecular dynamics calculations are of the same order of magnitude.[35] This is due to similar formation energies on Ta interstitials and oxygen vacancies at low oxygen potentials and similar heights of barriers for diffusion.[36] In order to keep the number of adjustable material parameters to a minimum, we have assumed that the diffusion coefficients of both tantalum and oxygen ions are the same. The expression for the diffusivity is:

$$D_i(T) = \frac{1}{2}\lambda^2 f e^{-\frac{E_{DIFF}}{kT}} \qquad (8)$$

where $\lambda$ is the effective hopping distance, *f* is the attempt frequency, and $E_{DIFF}$ is the activation energy for ion migration. The values of the attempt frequency and effective hopping distance are $10^{13} Hz$ and 0.032 nm, respectively.[37]

The flux due to temperature gradient has been considered only in few models of resistive switching and electroformation.[9,12,22,38] In solids, thermodiffusion has been studied mostly in $UO_{2-x}$ system due to composition changes observed in nuclear fuel rods.[39] In a macroscopic system as this, the typical temperature gradients are of the order of $10^3$ K/m. In nano-sized devices, they can be as high as $10^{10}$ K/m. The Soret coefficient multiplying the $c_i \nabla T$ term has a form $\frac{S_o^i}{kT^2}$ with $S_o^i$ representing the heat of transport of specie *i*. At this point, the heats of transport are known in very few materials systems and there is no established model predicting their values or signs. In the zero order approximation, one can replace heat of transport with the enthalpy of diffusion with sign "-" for vacancy- and "+" for interstitial-mediated diffusion.[40,41] This would suggest that the signs for Soret coefficients for oxygen and tantalum ions should be opposite. Given the above, we have considered several sets of heat of transport with $S_o^{Ta}$ value being positive or zero and $S_o^O$ either zero or negative.

The last term in the equation (6) considers the effects of stress during the electroforming process. In general, the thermo- and electro-migration terms tend to accumulate ions in certain portions of the structure and deplete them in others. For large composition changes, the corresponding strain energy is comparable to or can exceed other terms and its effects cannot be ignored. This term prevents unphysically large accumulation or depletion of ions. The absolute total concentration of ions at zero stress is taken as $6.5 \times 10^{22} atom/cm^3$. The initial concentration for Ta and O ions is $2.16 \times 10^{22} atom/cm^3$ and $4.33 \times atom/cm^3$, respectively.

2.2 Filament evolution under constant current

We have simulated two types of experiments described by Ma *et al*.[33] The first one used the current source applying a constant current pulse (200 $\mu A$) with the current ramping up over 1 ms.

The ramp time was much longer than the thermal time constant of the device allowing it to reach a quasi-steady-state temperature distribution throughout the process. To explore how each parameter influences the process of electroforming, we start with an oversimplified model assuming that there is no strain energy ($B = 0$), with $S_O^{Ta}$ = 0.05 eV, and $S_O^O$ = 0 eV. This eliminates second and third terms in equation (6) for oxygen flux. Since the initial oxygen distribution is uniform ($c_O$=const), the first term also vanishes, and the oxygen concentration stays constant everywhere throughout the process.

Figure 3 (a) shows the device voltage versus time with the inset showing the expanded view of the initial transient. During current ramp up, the device retraces the quasi-static *V-I* characteristic of a threshold switch: it reaches the maximum voltage of $V_{MAX}$=6.6 V at 0.06 ms and decreases to a quasi-steady-state value of 2.9 V for current of 200 µA. This behavior is induced by strong dependence of oxide conductivity on temperature and leads to decrease of voltage with increasing current and spontaneous current constriction.[30] At about 0.8 s, the device voltage drops rapidly again and stabilizes at about two thirds of the 2.9V with a long time constant of approximately 1.1 s. Since this relaxation is much longer than the device thermal time constant, the changes should be thought as due to ion redistribution.

Figure 3(b) shows the temperature profiles modelled at three different times marked in panel (a) as colored crosses. The vertical dashed black line indicates the active region of the device defined by the size of the top contact. The blue curve was calculated 20 ms after the beginning of the pulse at locations halfway through the oxide layer. This time is much longer than the thermal time constant of the device and corresponds to the temperature distribution after the threshold switch but before any significant ion motion has occurred. This statement is based on the blue trace of composition distribution in Figure 3 (c) which does not deviate from a flat line corresponding to initial uniform Ta fractional distribution of x=0.333. The highest temperature of 650K is reached at the center with the point of $\Delta T = \frac{T_{MAX}}{2}$ located 180 nm away. This is beyond the active region of the device. The temperature drop from the center to the edge of the active region is only 70 K.

As the time passes, the temperature gradient starts to drive the Ta ions towards the center of the device as the Soret coefficient for tantalum is positive. Initially the motion is slow due to low temperature (620 K) and low absolute radial gradient ($< 6.5 \times 10^8\ K/m$). As the center of the nascent filament becomes Ta-rich (red trace in Figure 3(c) at 800 ms)), it becomes more conductive, the temperature increases by 140 K (red trace in Figure 3(b)), and the voltage across the device starts to drop. This process forms a positive feedback loop in which composition changes enhance conductivity and cause $T_{MAX}$ and T gradient increase which, in turn, accelerate diffusion leading to a runaway and rapid decrease of resistance and voltage. Eventually the runaway slows down at about 1.5 s as the conductivity starts to saturate at high Ta content and the increasing concentration gradient results in significant Fick's diffusion. At the steady state, the maximum temperature within the filament reaches 900 K with the maximum gradient of $-1.3 \times 10^{10}\ K/m$. The Full Width at Half Maximum (FWHM) of the temperature distribution is 20 nm. The corresponding composition profile (black trace in Figure 3(c)) shows a maximum Ta content of 38.8% ($Ta_{0.39}O_{0.61}$) with the FWHM of 15 nm. In addition, it also shows a pronounced trench around the Ta-rich core with composition $Ta_{0.300}O_{0.700}$. Ta content is significantly lower here than the initial Ta density with material approaching composition of fully oxidized tantalum in $Ta_{0.286}O_{0.714}$. The trench forms mostly due to temperature distribution. If one considers a volume element at some distance from the center in initially uniform $Ta_xO_{1-x}$, the temperature and the diffusion rate at the face closer to the center is higher than at the outer face. The ion flux out of the element and toward the center is higher than the flux into the cell resulting in depletion of Ta. The rate of this process drops rapidly with distance from the center creating what appears as a steady state. The diameter of the trench, however, will continuously grow with time at temperature together with corresponding conductance of the device. This could represent the ultimate endurance limit for devices operating on this principle. The composition profiles shown in Figure 3(c) reflect main features found in the experimental Ta distribution. Ma et al.[16] reported the composition of the filament $Ta_xO_{1-x}$ to be x = 0.667-0.777, which is higher than the simulated value of x= 0.390. This deviation might be a result of overestimation of electrical conductivity in the simulation or Ta content in the filament in the experiment.

Figure 3 (d) shows the Ta ion density distribution at three different times. It has a similar shape as the Ta fraction x profile with the maximum Ta ion density at steady state of $2.78 \times 10^{22} atom/cm^3$. This value is about 27% higher than the initial Ta density. The corresponding increase of total ion density accounting for Ta and O ions is about 10% compared with the initial value. This implies unphysical level of stress much above yield strength of all materials in the structure. It indicates that the strain energy associated with the formation of the filament is very high and must be included in the equation (6) for realistic model of the composition evolution.

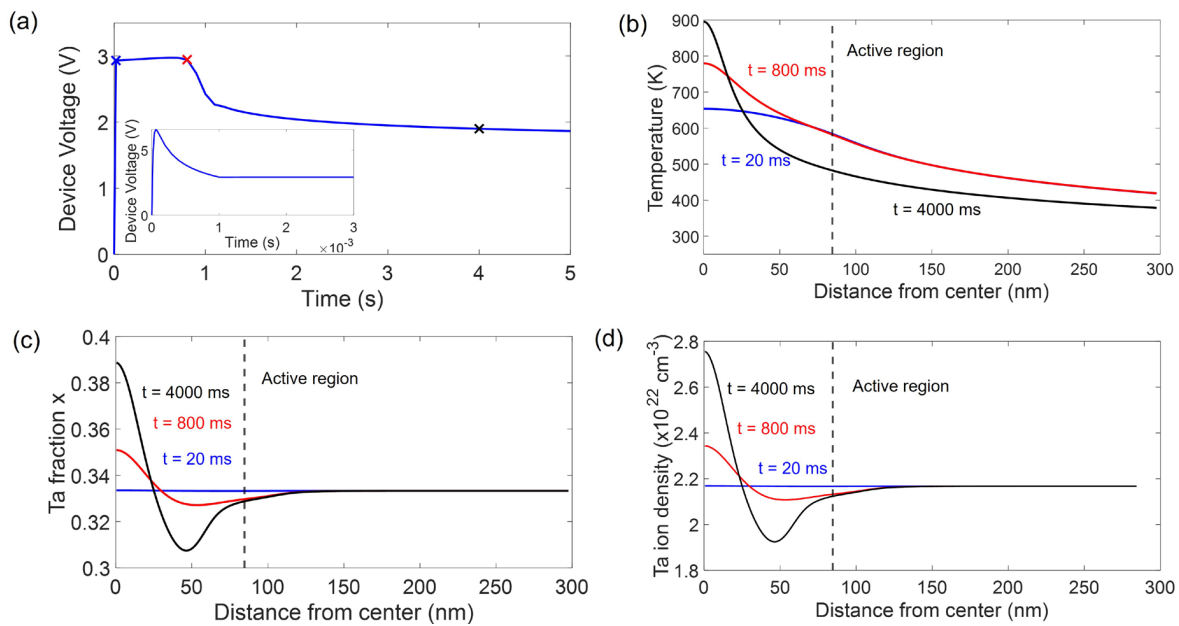

Figure 3 (a) Voltage across the device versus time for the current pulse of 200 $\mu A$ with 1 ms rise time for $S_o^{Ta}$ = 0.05 eV, $S_o^O$ = 0 eV, and $B$ = 0. The inset shows expanded view of the initial voltage transient with the overshoot caused by threshold switching. (b) Temperature distribution along the line in the middle of the functional layer at three different times. (c) Corresponding Ta fraction profiles. (d) Ta ion density profiles.

Stress term was included in the following simulation setting B=70 GPa while keeping all the other parameters the same as before including $S_o^{Ta}$ = 0.05 eV, and $S_o^O$ = 0 eV. We explore other values of bulk modulus in Supplementary Information (Figure S3). It is worth noting that there is no driving force for the segregation of oxygen other than the coupling with the tantalum motion through the stress term. Lastly, to make the results easy to compare, all the initial concentration values, boundary conditions, and current from the current source are the same. Figure 4(a) is the Ta density distribution at t = 20ms (before compositional runaway) and at t = 4000 ms at quasi-steady state. The density profile is similar to that without inclusion of the stress term: Ta accumulates in the center of the device while a ring of depleted Ta forms around it. The FWHM of the Ta distribution is 11 nm. The maximum Ta density is $2.68 \times 10^{22} atom/cm^3$ which is 3% lower than the simulation result without the stress. The depletion region has a diameter of 60 nm. The inner 40 nm of the depleted ring (from 20 to 60 nm) shows a deep depression followed by gradual Ta density increase. The minimum concentration of Ta is $1.93 \times 10^{22} atom/cm^3$ which is 11% lower than its initial value but almost the same as the simulated result in Figure 3. Figure 4 (b) shows the distribution of O ion density which forms almost a negative image of Ta distribution. Oxygen density in the center decreased by 11.5 % compared with initial density with the FWHM of depleted region of 12.0 nm. The removal of the oxygen ion from the core region relieves part of the strain energy generated by the tantalum ion accumulation. The oxygen density at the tantalum depleted ring increased by 5% compared to initial value. It is worth noting that the integrated quantity of either Ta or O in the cylinder with 80 nm radius is approximately the same as the initial value. Both types of ions redistribute within this volume and do not diffuse in or out of it. The total local ion density in the filament region remains almost the same with the changes below 0.3% (Figure 4 (c)). Another noticeable effect of adding the stress term is somewhat broader redistribution area compared to Figure 3 (d).

Figure 4 (d) shows the Ta fraction distribution along radial direction (the values of Ta fraction affect the electrical conductivity while the absolute densities do not). One can see overall shape of the curve closely following the one calculated while neglecting stress. The fraction of Ta in the center increased from 0.389 (without stress) to 0.412 (with stress). This represents 6% change. The conductivity change associated with such small Ta increase is still quite significant, as is visible in Figure 2 (a), increasing by a factor of 3. With increased conductivity in the filament region, more current will flow through this region heating up the device to higher temperature and generating higher temperature gradient. The balance between Fick's diffusion, stress induced drift, and thermodiffusion leads to FWHM of Ta fraction shrinking to 11 nm (by about 27%).

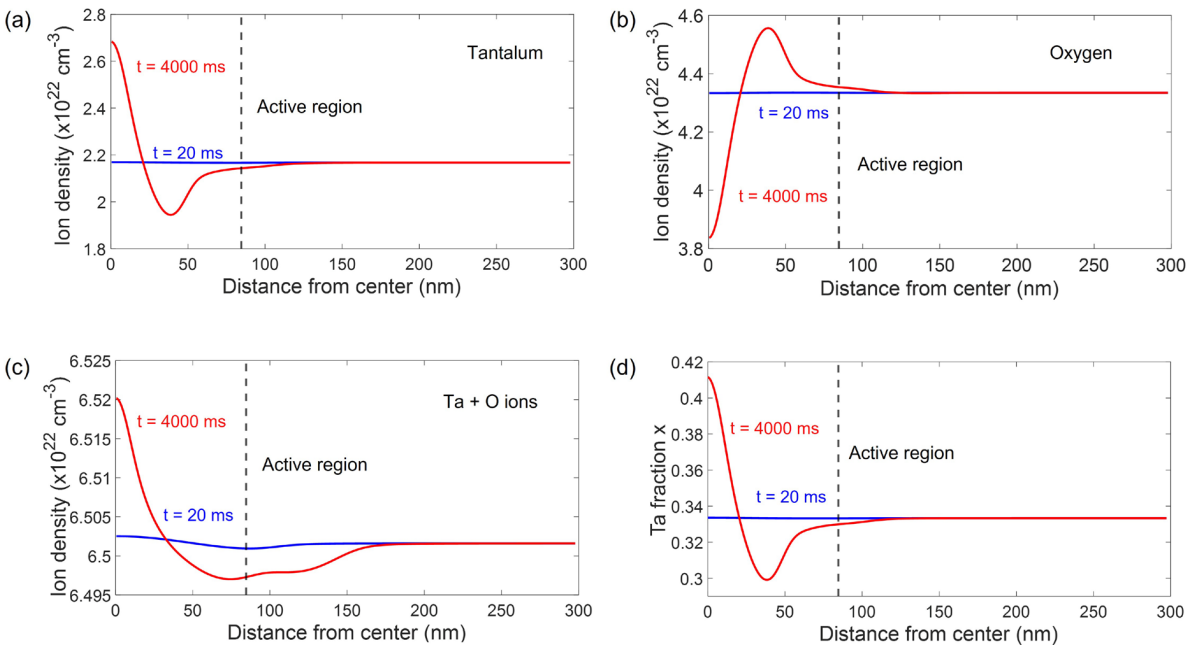

Figure 4 (a) and (b) Distribution of Ta and O ions before (blue line) and after electroforming (red curve) for the current pulse of 200 µA, $S_o^{Ta}$ = 0.05 eV, $S_o^O$ = 0 eV, and $B$ = 70 GPa. (c) Distribution of total density of ions. (d) Fractional composition of Ta before and after filament formation.

In summary, the effect of including the stress effects in the mass transport equation has not affected Ta distribution significantly for the same value of heat of transport. It induced oxygen, to move out of the area of Ta accumulation and move into the region where Ta was depleted to reduce strain energy of the system. Motion of oxygen increased the composition changes and conductance of the filament leading to decrease of the filament diameter.

One can easily extrapolate these results to cases of different values of $S_o^{Ta}$ and $S_o^O$ with high value of bulk modulus. For example, assuming that the heat of transport for oxygen is negative and $S_o^{Ta}$ = 0 would drive thermophobic O ions out of the center of device and produce oxygen-enriched ring around it. This would induce tensile and compressive stresses in areas of depletion and accumulation and force the Ta ions to move in direction opposite to that of oxygen. Resulting distribution will quite closely resemble the one shown in Figure 4 with detailed profiles shown in Supplementary Information (Figure S4). In essence, the elemental distributions are determined by the difference between heats of transport for the two elements. If $S_o^{Ta} > S_o^O > 0$, Ta will still form excess Ta in the center with oxygen depleted. The experimental elemental distributions, therefore, can be replicated by wide range of heats of transport as long as $S_o^{Ta} > S_o^O$.

Figure 5 shows the size of the filament as a function of $S_o^{Ta}$ with $S_o^O$=0 and $B$ = 70 GPa. The vertical axis is the FWHM of the Ta fraction x. The diameter of the filament is decreasing for increasing value of heat of transport with the size of the filament at $S_o^{Ta}$ = 0.05 eV being almost half of the value at $S_o^{Ta}$ = 0.03 eV. Also, the maximum density of Ta increases and the incubation time decreases with increasing value of S. For values of S above 0.1 eV the size of the filament becomes comparable to interatomic distances and the continuous model used here is no longer valid. For smaller values, the overall effect becomes weaker and the filament does not form.

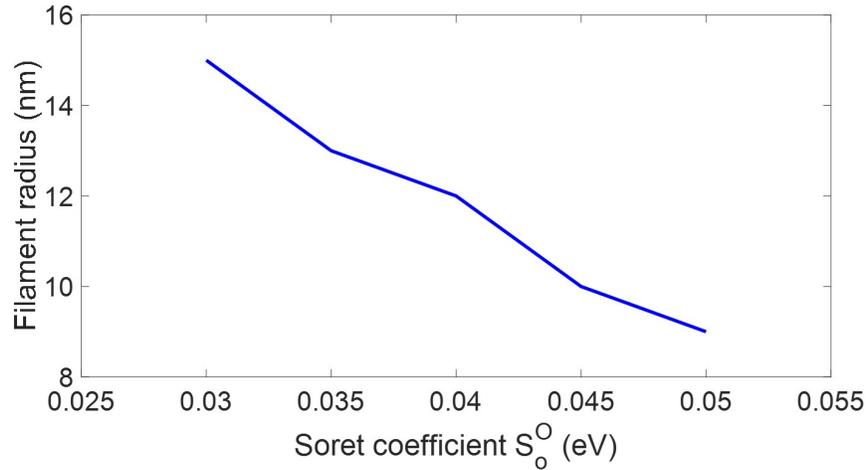

Figure 5 Filament size as a function of tantalum heat of transport $S_o^{Ta}$.

Ma *et al*. [33] has reported that the electroforming process is thermally activated with the time of electroforming determined by the temperature in the "hot zone". The activation energy was assumed to be equal to the activation energy for diffusion of more mobile ion (here, we assume the same activation energy for both ions in our simulation).

To test this assertion, we simulated the testing conditions used by Ma *et al*. Rather than using current pulse as was done in modelling leading to Figures 3-5, we use voltage pulse source connected to the device under test and a series load resistor (37 kOhm). Since the current source is equivalent to voltage source with infinite input resistance, this change is minimal as the load resistance is much higher than that of the device. It does not affect the ion distribution to any significant extent. The source voltage pulse had 1 $\mu s$ ramp up followed by a constant voltage lasting 50 s. The ramp time is comparable to the thermal time constant of the device. The blue, red, and black curves in Figure 6(a) show the current as a function of time with the source voltage equal 8V, 10V and 14V, respectively. All three curves show two abrupt current jumps. The first current jump is due to the threshold switching process. With the higher source voltage, the time to reach the threshold becomes shorter. It takes 1 $\mu s$ for the source voltage of 14 V device to reach its first plateau while 1.4 $\mu s$ for 10V source and 2 $\mu s$ for 8 V source. These times are clearly affected by the ramp up time, but they have been discussed extensively before[30] and of limited interest here. After threshold switching event, the temperature stabilizes for a relatively much longer period referred to as incubation time. These times are 17s, 2s and 0.02 s for 8, 10, and 14 V applied voltage, respectively. The explanation of the change is quite intuitive. Higher applied voltage leads to higher dissipated power density, and higher temperature within the device. This implies faster diffusion and, even for only slightly increased driving force of a temperature gradient, much shorter incubation time. The changes of composition and temperature are slow during incubation as evidenced by Figures 3-5 but the rate is continuously increasing with time leading to the second rapid increase of current.  This overall behavior agrees well with the experimental results.

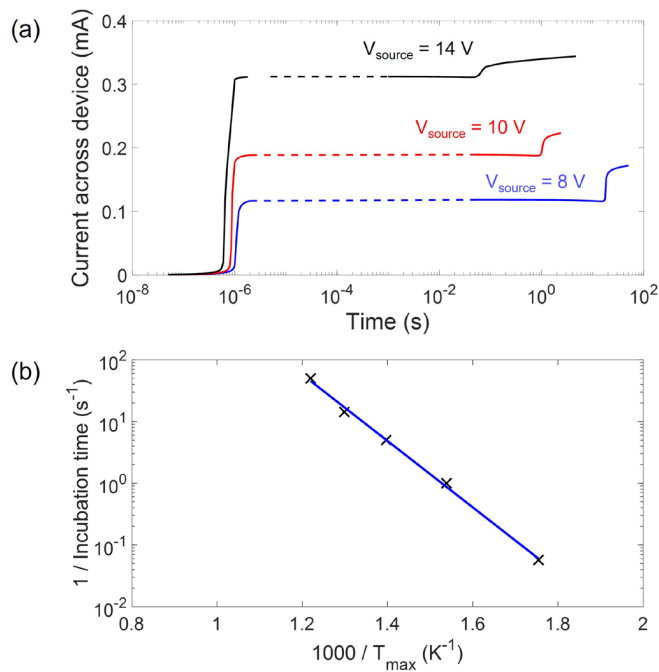

Figure 6 (a) Current across the device versus time with different applied source voltage. (b) Plot of inverse of incubation time versus temperature.

In Ma et al.'s work,[33] the change of the incubation time for compositional runaway with applied bias has been approximated by a thermally activated behavior by calculating the maximum temperature within the device after threshold switching for given value of bias. This assumed that the temperature during the incubation time is constant. However, Figure 3 (b) indicates that the temperature changes during the incubation time by about 100 K, clearly a not negligible amount. A better estimate is to use the average temperature during incubation. Figure 6 (b) shows the plot of the forming rate, defined as 1/ simulated incubation time, versus inverse of the maximum temperature in the film at half of incubation time. The plot shows a linear dependence corresponding to the activation energy of $E_a = 1.07\ eV$, which is close to our assumed value of the activation energy of diffusion $E_{DIFF} = 1.1\ eV$. This result validates the argument that the electroforming process is an Arrhenius process with the activation energy determined by the activation energy of thermal diffusion.

The model consisting of equations (1)-(8) with parameters used to create Figure. 5 and 6 was also used to simulate the commonly used way of electroforming by applying a quasi-static voltage sweep. The time for source voltage sweep was set to 1.2 s with the maximum voltage of 16 V with device and 37 kOhm load connected in series. It needs to be noted here that only a fraction of this voltage appears across the device at any point in time and the Figure 7(a) is plotted as a function of device voltage. The current initially increases super linearly and reaches 13 $\mu$A at the threshold voltage of 6.3 V. Soon after, the device experiences a thermal runaway rapidly transitioning along the load line denoted by the dashed part of the *I-V*. With the continuous increase of source voltage, the temperature of the device increases resulting in decreasing resistance. This causes the device voltage to drop as the device evolves along the continuous blue curve of the *I-V*. Despite this, the dissipated power and maximum temperature increase with source voltage and time. With increasing $T_{MAX}$, the radial temperature gradient driving Ta ions laterally increases as well eventually leading to the compositional runaway process marked by a second dashed line segment in Figure 7(a). The device at this point evolves with time even without change of source voltage. The process eventually slows down with the *I-V* turning up. On the return sweep of source voltage, the device does not retrace its original path as the changes taking part in the compositional runaway are permanent. This part of *I-V* is shown as red solid line. Somewhat surprisingly, the return path is not linear even though there is a cylindrical conductive filament connecting the two electrodes. To explain this behavior, we plotted the electrical conductivity as a function of position at three different times (Figure 7 (b)). The blue curve shows the temperature distribution at t = 0.05 s (marked as blue circle in panel (a)). As shown by the curve, the device is still at 300K and highly resistive everywhere. At the end of the upward sweep at t = 1.2 s (marked as red circle in panel (a)), the device is hot and has reached its quasi-steady state ion distribution and no more changes are expected on the way down. The corresponding conductivity is plotted as the red curve. At t = 2.4 s (black curve corresponding to the black circle in panel (a)), the applied voltage is 0 V, the temperature is back to 300K, and the conductivity everywhere in the device is lower than that at 1.2 s. This is to be expected as the conductivities in $Ta_xO_{1-x}$ increase with temperature. What is somewhat surprising is the magnitude of the increase. It is due to the small diameter of the filament what leads to high current densities and high temperatures. Moreover, very significant contribution is due to heating of the surrounding oxide. The radius of the cylinder with conductivity exceeding 5,000 S/m is 6.4 nm at 2.4 s but 16.5 nm at 1.2 s.

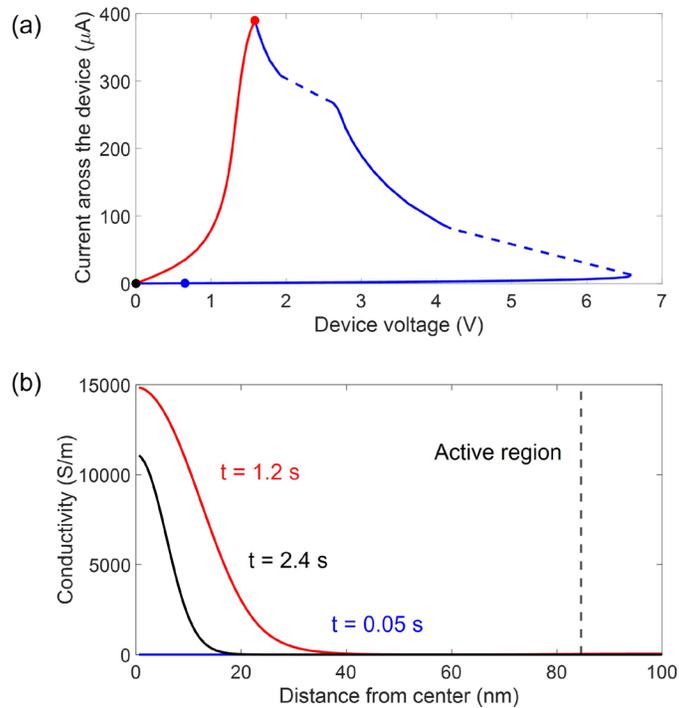

Figure 7 (a) Simulation of quasi-DC voltage sweep of a memory switching device. (b) Electrical conductivity as a function of position at three points (marked by circles) along the I-V.

Finally, one needs to address the question of whether the model presented above applies to other materials and device designs. We believe that one aspect of the model, the existence of two steps with the first one leading to temperature increase and second relying on ion motion, is almost universal. Ions in transition metal oxides such as $TaO_x$ and $HfO_x$ are generally not mobile at room temperature. Before rearrangement of ions can occur, the device needs to be heated to temperatures above 600 K. Since as-fabricated devices have low conductances, generating enough Joule heat requires high voltages not available in CMOS circuits. In the case considered here, the heating was made possible by the threshold switching event induced by thermal runaway. The net effect was a highly non-linear *I-V* of as fabricated device. While thermally induced threshold switching is expected in all systems that exhibit superlinear dependence of conductivity on temperature, the threshold voltages are too high for highly resistive oxides such as $HfO_2$. Highly non-linear *I-V* would have to originate in a different mechanism such as avalanching. Since virtually all experimental data indicate that the diameter of the filament is small, this mechanism would have to be local. It can be noted here that highly resistive oxides are typically used as very thin layers in switching devices. This results in electric field higher by at least an order of magnitude than one in structures considered here making local breakdown more likely. The nature of such localized current flow is not known.

The ion motion necessary in the second step of the process can either be driven by lateral temperature gradient, the process that is dominant here, or by electric field causing vertical redistribution of ions. It is quite apparent that vertical changes within the oxide itself cannot produce lower resistances typically reported after electroformation. Instead, local heating could allow for interdiffusion at the oxide-electrode interface and local loss of oxygen. This process was documented in $TiN/Ta_xO_{1-x}/TiN$ structures by Ma et al.[18] It can possibly be made more or less difficult by a proper selection of materials. Some authors suggest that the transfer of oxygen to electrode can be induced by electric field rather than concentration gradient. While in principle possible, such processes have not been confirmed or quantified experimentally.

In summary, the model presented here is fully consistent with a large set of electrical and structural data with mechanisms involved directly identified by experiments. It potentially applies to many device structures based on different materials. It is, however, not universal. It is likely that other structures operate based on different and yet unconfirmed mechanisms.

3. CONCLUSIONS

We have presented a finite element model of 150 nm TiN/Ta$_{0.33}$O$_{0.67}$/TiN resistive switching device responding to applied voltage and current pulses and undergoing electroformation. Model reproduced the accumulation of tantalum in the core of the filament and its depletion in the surrounding cylinder with the oxygen concentration decreasing in the core and increasing at the outer periphery of the filament. This is due to thermodiffusion term in expression for mass transport with Ta being thermophilic ($S_o^{Ta} > 0$) and/or O being thermophobic ($S_o^O < 0$). To replicate the experimental ion distribution, it is enough for one of the heats of transport to be different from zero with stress coupling responsible for motion of the other type of ion. The size of the core is sensitively dependent on the value of S-parameters decreasing with increase of S.

4. METHODS

The Ta$_x$O$_{1-x}$ films were deposit using reactive sputtering method with chamber pressure set to 3 mTorr and the power at 50 W. A mixture of Ar and O gas were flow into the system as the reactive gas. The composition of Ta$_x$O$_{1-x}$ was controlled by adjusting the oxygen and argon flow rates. The oxygen content with respect to oxygen and argon flow rates was measured by Atom Probe Tomography (APT). Needle shaped specimens were fabricated from layered Ta$_x$O$_{1-x}$ films using standard focused ion beam liftout and milling techniques described by Thompson *et al.* [42] The APT experiments were run in a CAMECA LEAP 4000XHR with a 60 pJ laser power, 100 kHz pulse repetition rate, 50K base temperature, and a 0.2% detection rate. The experimental results and regression fit are shown in Figure S2 in Supplementary Information. Electrical conductivity was measured using circular transmission line model structure and the hot stage.

Modelling was performed using a commercial Comsol software package. All material properties are listed in Table S1 in Supplementary Information. Voltage was applied to the upper surface of the top electrode with lower surface of the bottom electrode grounded. The entire structure was placed on 500 nm thick $SiO_2$ with the bottom surface at 300 K and all other surfaces thermally insulated.

ACKNOWLEDGMENTS

This work was supported by the National Science Foundation (NSF) under Grant No. DMR-1905648. The authors acknowledge the use of the Materials Characterization Facility at Carnegie Mellon University supported by grant MCF-677785. APT was conducted at ORNL's Center for Nanophase Materials Sciences (CNMS), which is a U.S. DOE Office of Science User Facility.

SUPPLEMENTARY INFORMATION

# Simulation of lateral ion migration during electroforming process

Jingjia Meng[1], Enkui Lian[1], Jonathan D. Poplawsky[2], Marek Skowronski[1]

[1]Department of Materials Science and Engineering, Carnegie Mellon University, Pittsburgh, Pennsylvania, 15213, USA

[2] Center for Nanophase Materials Sciences, Oak Ridge National Laboratory, Oak Ridge, Tennessee 37831, United States

## S1. Device fabrication

Switching device modelled in this work was fabricated by sputtering of 40 nm of TiN on oxidized silicon wafer and patterning of bottom electrode and the load resistor serpentine by ion milling. The functional oxide was reactively sputtered at flowrate of Ar:O = 60:2.7 sccm, power of 50 W, and chamber pressure of 3 mTorr. The $SiO_2$ layer was deposited on top of $Ta_xO_{1-x}$ layer without breaking vacuum and patterned with e-beam lithography and Reactive Ion Etching. The top electrode was deposited at the same condition as bottom electrode and patterned by liftoff.

## S2. Electrical conductivity structures and measurement

50 nm thick $Ta_xO_{1-x}$ layers of different compositions were sputtered at different oxygen partial pressures by changing the Ar:O ratio during reactive sputtering process. The films were deposited at a fixed chamber pressure of 3 mTorr and a power of 50W. The film composition was determined by Atom Probe Tomography with results shown in Figure S1. The blue crosses represent the experimental points with regression fit shown as red line. Function expressing the Ta fraction x as a function of oxygen flow rate is given by:

$$x_{Ta} = \left(-1.1 \times fl + \frac{2.19}{fl} + 33.155\right) \times 10^{-2}$$

where fl is the flow rate of oxygen (the flow rate of Ar is fixed at 60 sccm). Circular transmission line model structures were formed $Ta_xO_{1-x}$ layers by sputtering 20 nm thick TiN followed by 100 nm of gold. Structures were patterned by photolithography and liftoff.

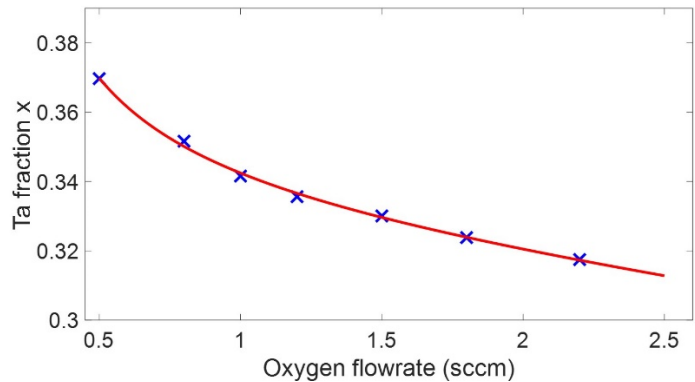

## S3. COMSOL simulation

Figure S2 shows the geometry of the device in the simulation. The voltage source was connected to the upper surface of the top TiN electrode marked by a blue line while the lower surface of the bottom electrode is grounded. The red dashed line marks the device periphery that was electrically insulated. The device was part of a larger slab used for thermal simulation. The radius of the slab was 10 μm with the thickness of 11.13 μm. The slab ended with $SiO_2$ layer neglecting the underlying silicon. The bottom surface of the slab was set to 300K with all other surfaces thermally insulated. Table S1 lists all materials' properties used in the simulation. Volumes of Ta and O ions were set to $1.99 \times 10^{-30}$ m³ and $8.37 \times 10^{-30}$ m³, respectively.

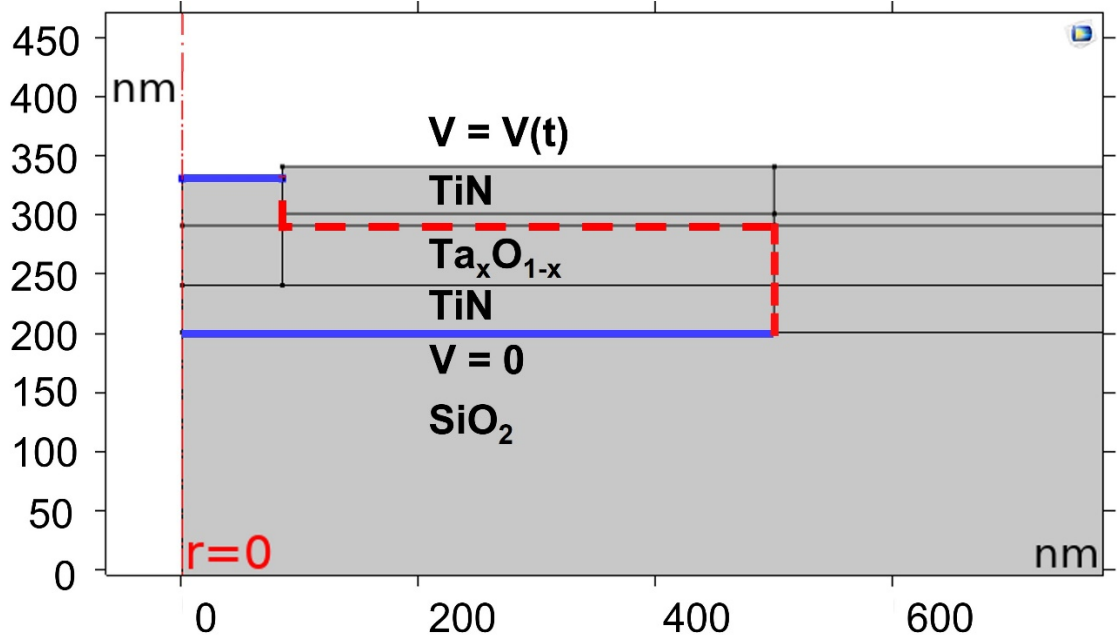

Table S1. Material Properties

| Materials | Thermal Conductivity $W \cdot m^{-1} \cdot K^{-1}$ | Heat capacity $J \cdot kg^{-1} \cdot K^{-1}$ | Density $kg \cdot m^3$ | Relative Permittivity |
|---|---|---|---|---|
| Si | 130 | 700 | 2329 | NA |
| SiO$_2$ | 1.4 | 730 | 2200 | NA |
| TiN | 5 | 545 | 5210 | 4 |
| Ta$_x$O$_{1-x}$ | 0.6 | 174 | 8200 | 22 |

## S4. Additional simulation results

As discussed in the main text, we found that the size of the filament is sensitive to the value of heat of transport and conductivity of the functional oxide. It is not sensitive to the value of bulk modulus as apparent in Figure S3 (a). However, as shown in Figure S3 (b), the total ion density decreases with increase of B.

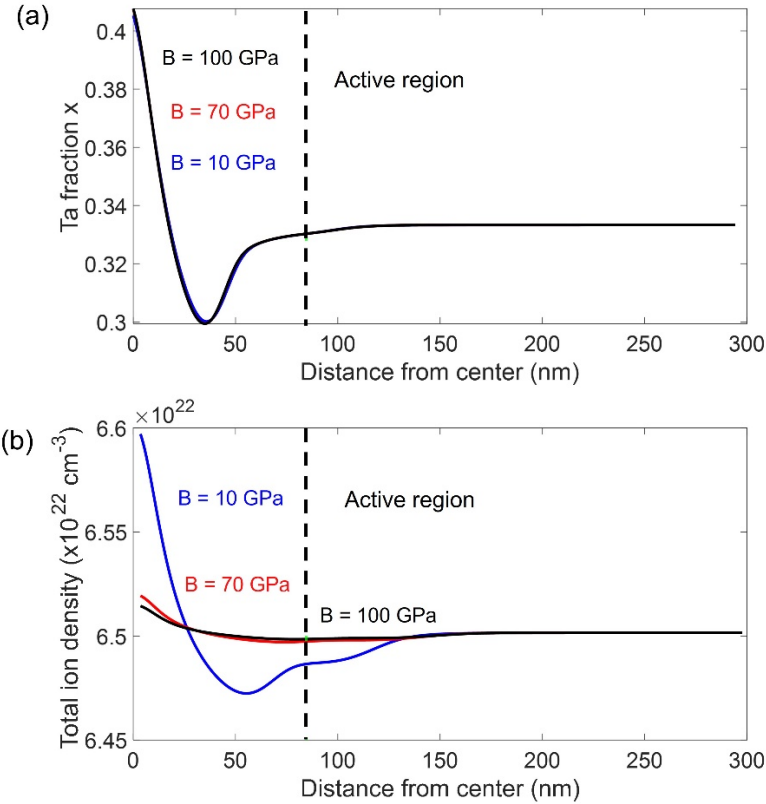

Figure S4 shows the distributions of Ta and O ions in the device for negative value of oxygen heat of transport ( $S_o^O$ = -0.1 eV) with $S_o^{Ta}$ = 0 and B=70 GPa. Both distributions are similar to these in Figure 4 (a) and (b). Oxygen is moving out of the center, as a result of thermodiffusion, forming a depleted region at the center and O-enriched ring around it. The O density at the center is 8% lower than the initial value while at the ring, the density is 3.9% higher than the initial value. The FWHM of the central minimum is 16.8 nm. In response to oxygen motion, tantalum accumulates in the center and is depleted in the ring. The maximum Ta ion density change at the center region is 13.6% which is about 12% lower than that in Figure 4 (a). The overall ion distributions show the same shape as the case of $S_o^{Ta}$ = 0.05 eV and $S_o^O$ = 0 discussed in the main text.

Different from Figure 4 (c), the total ion density in the center is lower than the initial value by 1% and the affected region (ion redistribution region) is around 200 nm. Also, the tantalum fraction x is only 0.38 which is 7% lower than that in Figure 4 (d). It is worth noticing that the heat of transport driving the segregation is higher in this simulation compared to ones in the main text (lower than this value will not form the filament in a reasonable time).

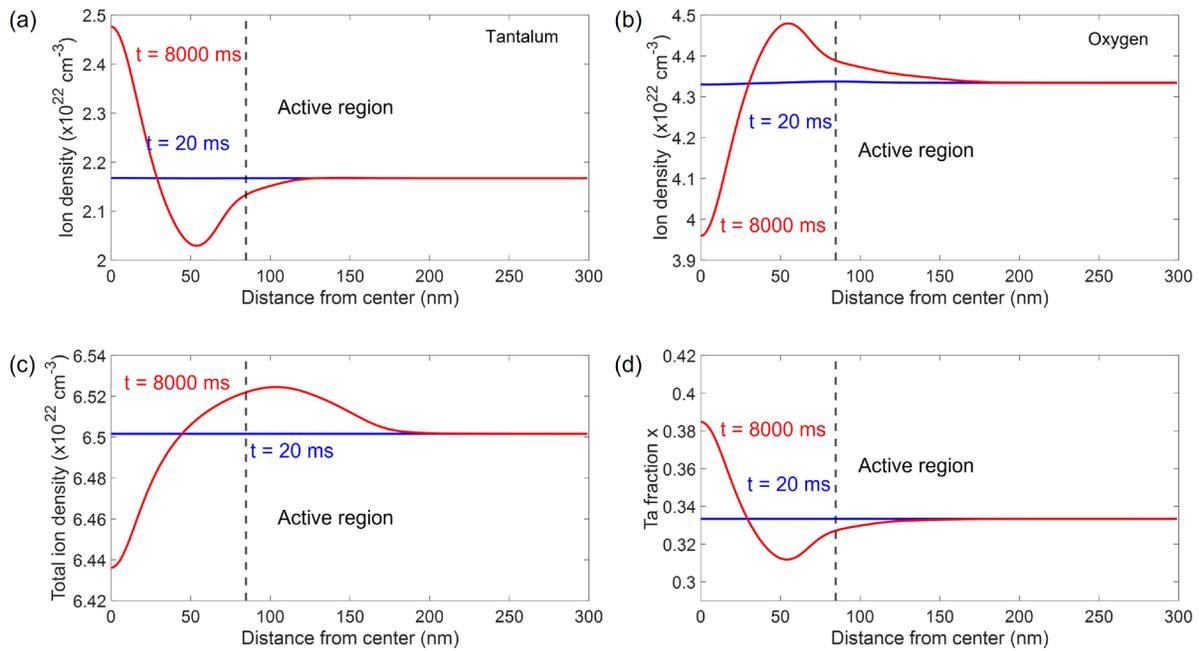

Figure S4. (a) and (b) Distribution of Ta and O ions before (blue line) and after electroforming (red curve) for the current pulse of 200 $\mu$A and $S_o^O$ = -0.1 eV, $S_o^{Ta}$ = 0, B=70 GPa. (c) Distribution of total density of ions. (d) Fractional composition of Ta before and after filament formation.

The last case to consider is for non-zero heats of transport for both oxygen ($S_o^O$ = -0.01 eV) and tantalum ($S_o^{Ta}$ = 0.04 eV) and stress B=70 GPa. As before, all other parameters remain the same. Figure S5 (a) shows the oxygen ion density profile with almost identical distribution as one in Figure 4(b). The oxygen concentration in the center induced by negative value of its heat of transport decreased by less than 2%. The total ion density profile in Figure S4(b) shows 2% decrease in the center compared with Figure 4 (c).

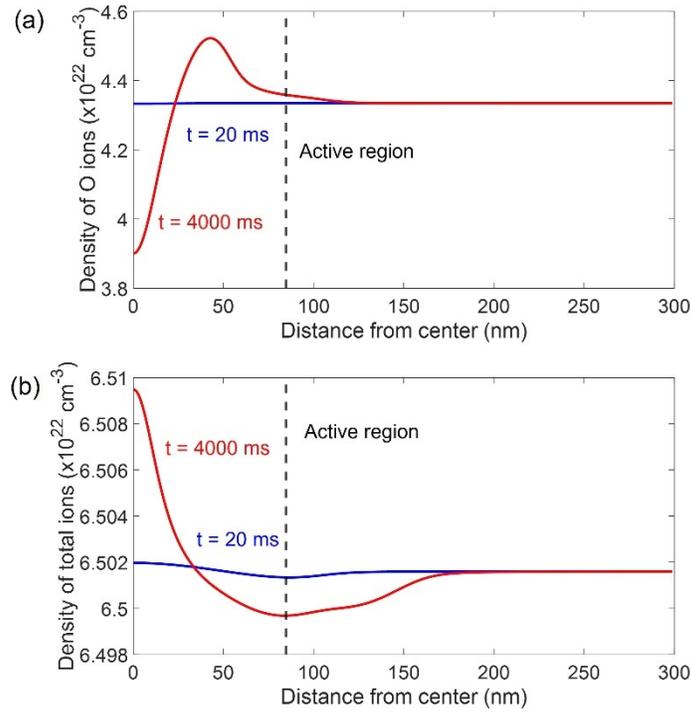

Figure S5 (a) Distribution of O ions before (blue line) and after electroforming (red curve) for negative heat of transport of O under the current pulse of 200 $\mu$A and $S_o^O$ = -0.01 eV, $S_o^{Ta}$ = 0.04 eV, B=70 GPa. (b) Distribution of total density of ions.